\documentclass[aip,jcp,preprint]{revtex4-1}
\usepackage{graphicx}
\usepackage{latexsym}
\usepackage{amsmath}
\usepackage{amssymb}
\begin{document}
\title{Phase transformations in methanol at high pressure measured by dielectric spectroscopy technique.}
\author{M.V.Kondrin}
\email{mkondrin@hppi.troitsk.ru}
\affiliation{Institute for High Pressure Physics RAS, 142190 Troitsk, Moscow, Russia}
\author{A.A. Pronin}
\affiliation{General Physics Institute RAS, 117942 Moscow, Russia}
\author{Y.B. Lebed}
\affiliation{Institute for Nuclear Research RAS, 142190 Moscow, Russia}
\author{V.V. Brazhkin}
\affiliation{Institute for High Pressure Physics RAS, 142190 Troitsk, Moscow, Russia}
\begin{abstract}
Dielectric response in methanol measured in wide pressure and temperature range ($P < 6.0$ GPa; 100 K $<T<$ 360 K) reveals a series of anomalies which can be interpreted as a transformation between several solid phases of methanol including a hitherto unknown high-pressure low-temperature phase with stability range $P > $ 1.2 GPa and $T<270$ K. In the intermediate P-T region $P \approx 3.4-3.7$ GPa, $T \approx 260-280$ K a set of complicated structural transformations occurs involving four methanol crystalline structures. At higher pressures within a narrow range $P \approx 4.3-4.5$ GPa methanol can be obtained in the form of fragile glass ($T_g \approx 200$ K, $m_p \approx 80$ at $P= 4.5$ GPa) by relatively slow cooling.   
\end{abstract}
\maketitle
\section{Dielectric spectroscopy and phase diagram of methanol}
Methanol is a very interesting object for study of its P-T phase diagram because it is the simplest organic substance with only one hydrogen bond per molecule. It can be regarded as a simple (one bonded) approximation of water where one of the hydrogen bonds is capped with alcyl group. It is well-known that the P-T phase diagram of water is extremely complicated, but it turns out that by now it is better studied than the phase diagram of its significantly simpler interconnected counterpart -- methanol.

At ambient pressure there are two crystalline phases of methanol -- the high-temperature one (just below the melting curve) is a plastic crystal, orientationally disordered  $\beta$-phase which exists in the temperature range $T=169-155$ K. At lower temperatures $\beta$-phase transforms into an ordered $\alpha$-phase, stable down to lowest temperatures. It was shown that both these phases persist at least up to pressures about 1.6 GPa \cite{gromnitskaya:jetpl04}. There is also one report \cite{allan:prb98} where it was demonstrated by the single crystal X-Ray diffraction in diamond anvils, that the high-pressure phase of methanol stable at room temperature in the pressure range $4-6$ GPa (called $\gamma$-phase) is different from the $\alpha$-phase. It should be noted, that this observation contradicts earlier optical studies at high pressures \cite{mammone:jpc80,eaton:cpl97}, where no difference between high- and low-pressure phases was found.

However the phase boundaries, not only between the crystalline phases, but even the melting curve above 2 GPa are not known yet (though the efforts of its determination \cite{bridgmen:paaas42} date back to Bridgman himself). The low entropy and volume changes accompanying the transitions ($\Delta V_{\alpha-\beta}/V_0 =$ 0.6 \% , $\Delta V_{melt}/V_{\beta} =$ 3 \% \cite{staveley:jcs54}, $\Delta S_{\alpha-\beta} = $ 3.8, $\Delta S_{melt}=$ 18.0 J/(K mole) \cite{carlson:jcp71} at ambient pressure) require high sensitivity of standard methods based on measurement of energy output (like DTA/DSC technique, though these sort of methods were also tried at elevated pressures \cite{wurflinger:jpcs77}).  Another complication of experimental studies of methanol at high pressure is the easiness with which methanol can be obtained in metastable liquid phase (supercooled or superpressed) -- the feature marked by many researchers who studied methanol at high pressures.  For example, the $P-T$ region, for which the viscosity data for liquid methanol \cite{grocholski:jcp05,cook-herbst:jpc93} were obtained,  overlaps with the region of existence of solid $\gamma$-phase \cite{allan:prb98}. Computer simulation \cite{gonzales:jcp10} of methanol phase diagram (despite the tremendous progress made by the computational molecular dynamic in the recent years) predicted hugely overestimated values for transition temperatures and pressures. Though at ambient pressure methanol is practically impossible to obtain in glassy form by the cooling of liquid \cite{anderson:jrs88}, the existence of methanol glass at elevated pressures was widely discussed \cite{piermarini:jap73,brugmans:jcp95,zaug:jpc94,grocholski:jcp05,cook-herbst:jpc93}, so methanol is an interesting model object  for study of the transitions between various (and variously) disordered phases at high pressure. Structure determination of material composed of light atoms by direct diffraction methods is very technically complicated, especially at high pressures \cite{loveday:prl00,loveday:n01}.  So indirect methods of location of possible phase boundaries (which could be lately refined by the structural studies) are very important in this case.

Liquid and solid methanol at ambient pressure was thoroughly studied by the dielectric spectroscopy (DS) technique \cite{denney:jcp55,davidson:cjc57,ledwig:zpc82,barthel:cpl90}. This studies give consistent picture of phase transformations in methanol at ambient pressure. Since methanol molecules are highly polar, any arrest of their motions should invariably result in significant change of dielectric susceptibility values. Just that is observed in practice in a sequence of transitions from liquid to $\beta$-phase and then into $\alpha$-phase ($\varepsilon \approx 70, 6, 3$ respectively \cite{denney:jcp55,davidson:cjc57}). These observations contributed a lot to understanding of disordered nature of $\beta$-phase. Due to its independence of scanning speed (in contrast to DTA/DSC one) DS technique also allows to avoid the problem of metastable phases and to  set practically arbitrary routine for heating/cooling cycles. Combination of all these virtues makes DS almost the method of choice for studies of phase diagram of polar compounds at high pressure. So far, due to the characteristics of standard cylinder-piston setup, the DS applicability was limited to pressures $P<1.8$ GPa. Only recently this range was significantly extended \cite{pronin:pre10,pronin:jetpl10,kondrin:jcp12} by introduction of toroid-type high-pressure device \cite{hpr:khvostantsev2004}. In this paper we present results of DS of methanol for pressure and temperature region P$=0-6$ GPa, T$=100-360$ K  and discuss the consequences this data imply for determination of crystal structure of solid methanol in this range. 

\section{Experiment details}   

The samples of 99.5 \% pure methanol (MERCK) were used. The previous research \cite{wurflinger:jpcs77} demonstrated that the influence of impurities ($<$ 1 at. \%) on phase transformations in methanol is negligible. In general we follow the same experimental routine as it was outlined earlier \cite{pronin:pre10,pronin:jetpl10,kondrin:jcp12}. Experimentally accessible frequency window (10 Hz -- 2 MHz) and the precision of dielectric susceptibility measurements ($\Delta \varepsilon \approx 0.1$) was determined by the device used -- QuadTech-7600\cite{quadtech:7600}.  The capacitance values were measured twice before an experiment: in empty capacitor and capacitor filled with methanol. These values were used later for calculation of dielectric susceptibility at high pressure. ``Empty'' capacitance readings were about 10 pF, values of ``filled'' capacitance were related to it according to  $\varepsilon \approx$ 30 characteristic of methanol at room temperature.  During high-pressure experiments the variation of thermodynamic parameters in crystal phase was possible only along isobars (with typical rate $\pm 0.5$ K/min), because otherwise the pressure change would lead to the breakage of the measurement capacitor, so the presented data were exclusively isobaric ones. However the small deformation of measurement capacitor can be expected even in this ``mild'' regime but the checks of capacitance values after pressure release demonstrated that such deformations were negligible. It was found that the final values of ``empty'' capacitance are within 5\% range of the starting ones. The pressure values during experiments was estimated by the readings of manganin gauge in the liquid phase  just prior to the crystallization onset. The rate of temperature variation was not strictly controlled, mostly  it was determined by the thermal inertia (quite large) of our setup. The typical cooling rate was about  $-0.5$ K/min but the heating rate can be varied in the range $0.5 - 1$ K/min by the application of external manual heating. Though we weren't able to repeat exactly cooling/heating cycle at the same P-T conditions  several times in a row, but subsequent measurements in similar P-T conditions produced similar results. Since the heating rate was quite small (in comparison, for example, to the ``standard'' value of 10 K/min adopted for measurements of glass transitions in DTA/DSC experiments) we didn't observe significant difference of phase transitions temperatures (from the values reported below) depending on  the temperature scan rates. Presented results were collected in the process of  several  high-pressure experiments on different measurement cells. 

\section{Dielectric spectroscopy of solid methanol at high pressure}
\begin{figure}
\centering
\includegraphics[width=0.9\columnwidth]{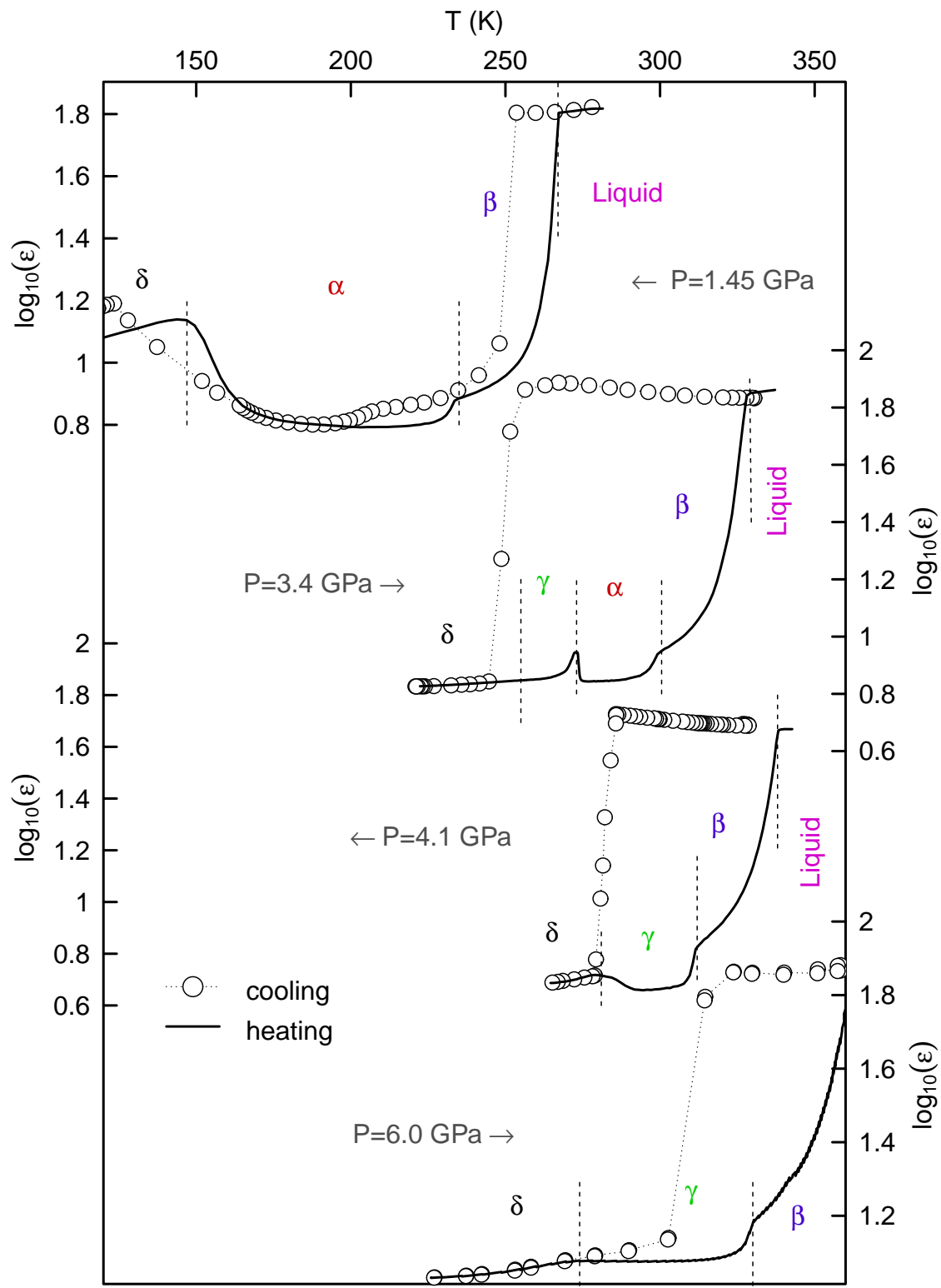}
\caption{Temperature scans of static dielectric susceptibility collected at different pressures on the frequency $\log_{10}(\nu/\text{Hz})=5.4$. Vertical dashed lines designate the phase boundaries. }
\label{eps}
\end{figure}

\begin{figure}
\centering
\includegraphics[angle=270,width=0.9\columnwidth,clip=F]{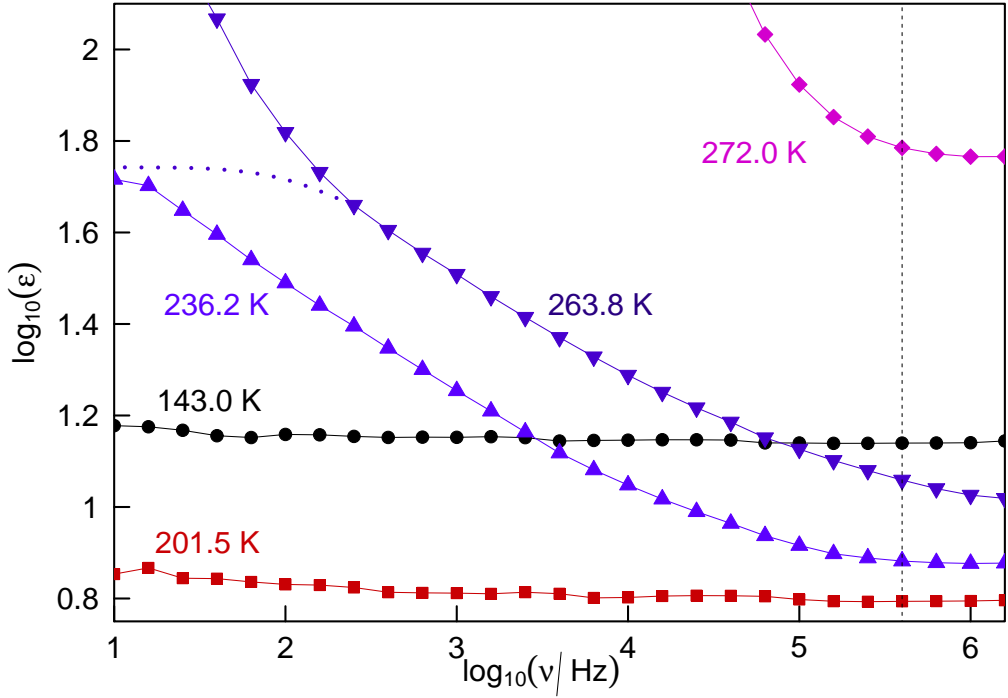}
\caption{ Dispersion of dielectric susceptibility in different phases of methanol at pressure $P=$ 1.46 GPa. The measurement temperatures are shown beside corresponding curves. Significant dispersion is observed in disordered methanol phases (liquid -- $\blacklozenge$  and plastic crystalline $\beta$ -- $\blacktriangle$, $\blacktriangledown$) and is absent in the ordered ones ($\alpha$ -- $\blacksquare$, $\delta$ -- $\bullet$  and not shown here $\gamma$-phase, see Fig.~\ref{ag} too). Vertical dashed line indicates the measurement frequency $\nu_{0} \approx $ 300 kHz.}
\label{disp}
\end{figure}

\begin{figure}
\centering 
\includegraphics[angle=270,width=0.9\columnwidth]{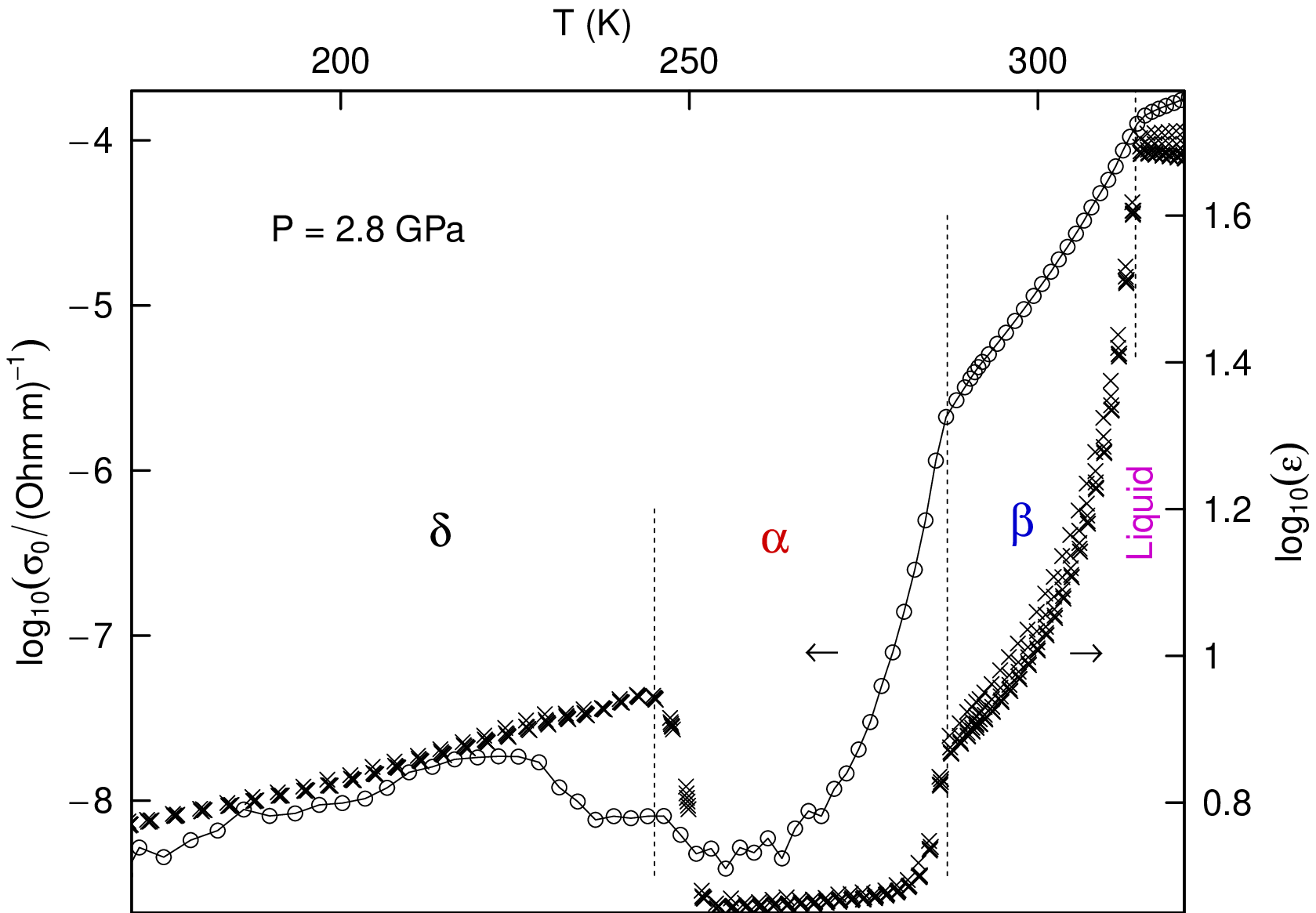} 
\includegraphics[angle=270,width=0.9\columnwidth]{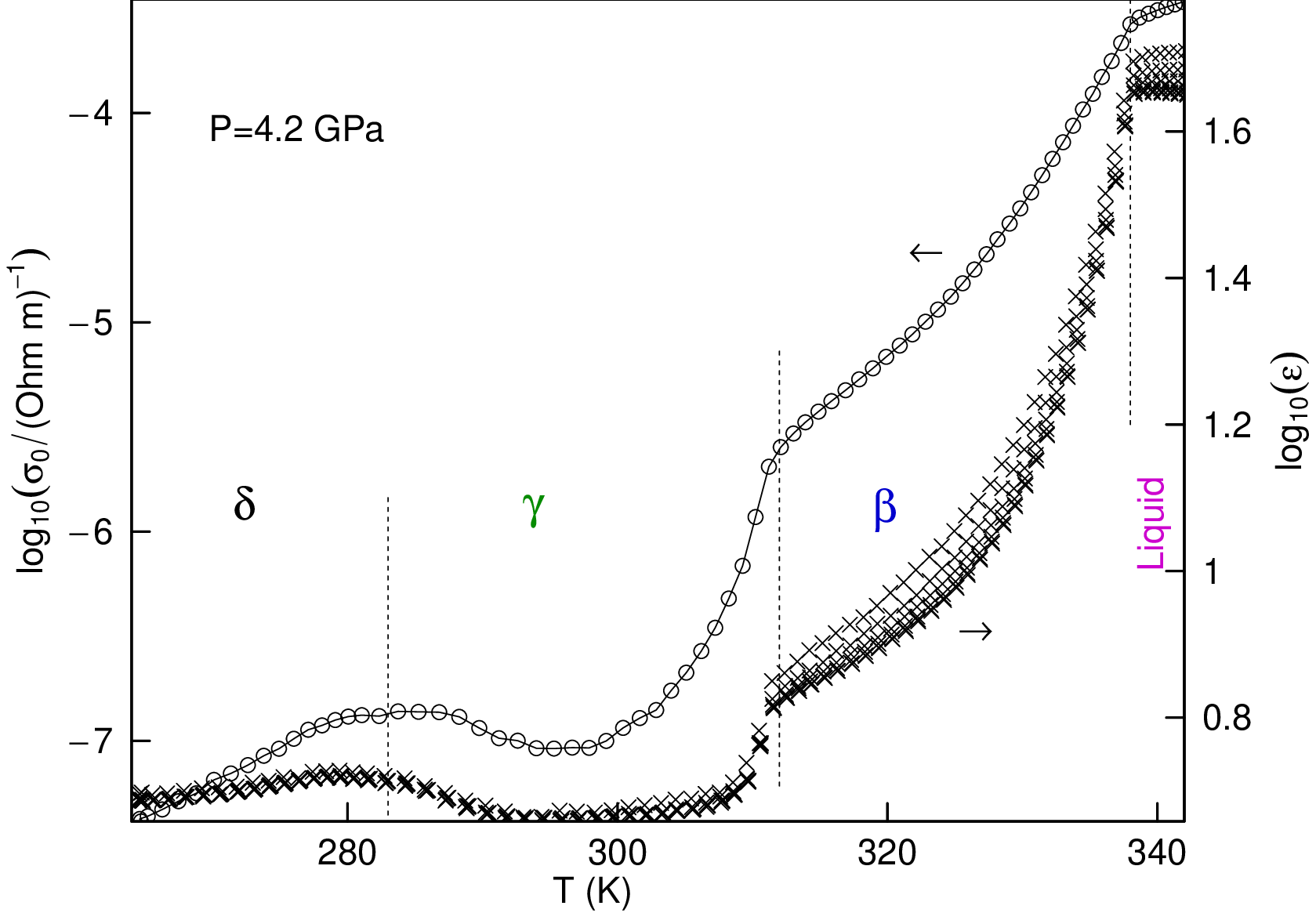} 
\caption{The temperature scans of static conductivity $\sigma_0$ ($\circ$) and high-frequency dielectric susceptibility $\varepsilon(\nu)$ (frequency range $\nu=$ 100 kHz -- 1 MHz, $\times$) collected  along two isobars just before and after $\alpha - \gamma$ transformation. Vertical dashed lines designate  the phase boundaries.}
\label{ag}
\end{figure}

The temperature dependencies of dielectric susceptibility values measured at frequency $\nu_0 \approx$ 300 kHz at different pressures are shown in Fig.~\ref{eps}. The choice of the measurement frequency was influenced by the presence of dispersion observed in our frequency window in unordered methanol phases (some examples at $P=1.45$ GPa are shown in Figs.~\ref{disp}). Low-frequency dispersion commonly found in liquids is a surface effect and  is usually attributed to contact polarization. On the other hand,  dielectric dispersion in plastic crystals is an intrinsic/volume property and it is widely observed in plastic crystal phases of mono-alcohols (for example, metastable phase of ethanol \cite{benkhof:jpcm98} or orientationally disordered phases of cyclo-octanol/heptanol \cite{martinez:jncs11}), so methanol is not an exception. Though the dispersion induced by polarization may occur in plastic crystal phases too, especially at high temperatures  and the low-frequency range (see the difference between experimentally registered values at $T=263.8$ K and the dotted line representing the polarization-less dispersion in Fig.~\ref{disp}), but this is not the most prominent feature. Though it may seem quite counter-intuitive, the ``static'' dielectric susceptibility isn't the lower-frequency limit  but the frequency-independent value above dispersion region (in the particular case of $\beta$-methanol we just follow the earlier practice \cite{denney:jcp55,davidson:cjc57}). The frequency value $\nu_0 \approx$ 300 kHz was above dispersion region in the liquid as well as in the $\beta$-methanol phase so the dielectric susceptibility value measured at this frequency was considered as a ``static'' one. In all other ordered methanol phases  dispersion wasn't observed so the values measured in such a way practically coincide with the low-frequency ones (Figs.~\ref{disp},~\ref{ag}).

\begin{figure} 
\centering 
\includegraphics[angle=270,width=0.9\columnwidth]{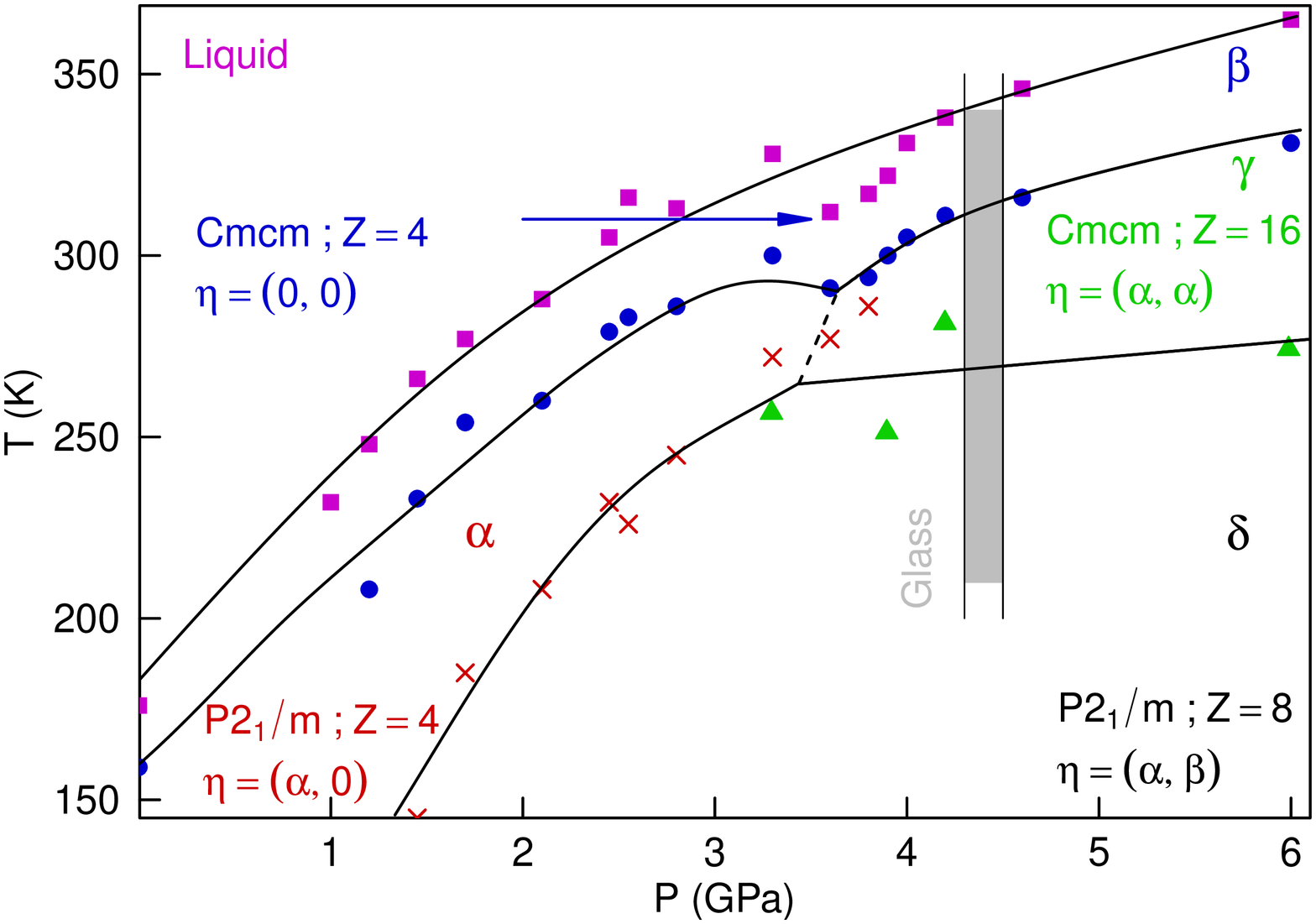} 
\caption{The phase diagram of methanol based on  anomalies of static dielectric constant from Fig.~\ref{eps}. Points correspond to experimental data and solid lines are just guides to eyes. The labels designate the solid phases of methanol. Grey area indicates the region of methanol glass formation.} 
\label{pt} 
\end{figure}

As it can be seen from Fig.~\ref{eps}, the hysteresis between cooling  and heating curves spans quite a wide temperature range which significantly increases above 2.0 GPa and exceeds 50 K at $P=6$ GPa. However at all pressures used, the crystallization of the liquid phase did occur as  indicated by abrupt drops of dielectric response from 70 (values typical to liquid methanol phase at melting point) to several units. In the process of subsequent heating it was revealed that the supercooled region includes several features (marked by vertical dashed lines in Fig.~\ref{eps}) which should be interpreted as  the structural transformations of solid methanol between different crystalline phases. The whole set of these anomalies plotted on P-T plane produces consistent picture (Fig.~\ref{pt}) which is considered as a close approximation of the true thermodynamic phase diagram of methanol. 

Though connection between the anomalies in temperature dependence of static dielectric susceptibility and structural transformation is not that straightforward, it is widely observed in practice. Static dielectric susceptibility in quite a number of polar molecular compounds is remarkably robust to variation of temperature. In disordered materials (like liquids) it is inversely proportional to temperature according to the Debye-Onsager-Kirkwood law and is independent of temperature in most of crystalline solids, so in that last case it  well deserves its another name "dielectric constant". In particular, abrupt changes (especially exponentially large) of dielectric susceptibility value is a good indication of phase transformation taking place in the material studied. 

Comparison of this tentative phase diagram (Fig.~\ref{pt}) with available diffraction data \cite{tauer:ac52,torrie:mp89,torrie:jssc02,allan:prb98} suggests existence of two high-pressure phases of methanol: the room-temperature one (in pressure range coinciding with that of $\gamma$-phase reported previously \cite{allan:prb98}) and unknown before low-temperature phase (marked as $\delta$-phase in Fig.~\ref{pt}). The transformations of this phase to higher temperatures ones is accompanied by quite significant increase of dielectric constant, though at high pressures (curve P=6.0 GPa in Fig.~\ref{eps}) this transition may be somewhat smoothed due to nonhydrostatic effects which we can not rule out in our experiments at such a high pressure. Determination of the lowest-pressure extreme point of $\alpha-\delta$ phase boundary was limited by the experimentally available temperature range but at $T=$100 K  this boundary is found to be in the pressure range $P=1.2 -1.4$ GPa. The  $\delta$-phase seems to be more conductive than the $\alpha$ and $\gamma$ phases of methanol as suggested by conductivity data in Fig.~\ref{ag}. However we can not bring forward any quantitative assessments, because the conductivity level is very low in this case (the resistance of $\delta$ phase at 2.8 GPa is about of 10~GOhm~cm, which is on the edge of capabilities of measurement device\cite{quadtech:7600}), so the conductivity values below $10 ^ {-8}$ (Ohm~m)$^{-1}$ in Fig.~\ref{ag} should be considered with certain caution.   

Another conclusion following from the phase diagram Fig.~\ref{pt} is the presence of disordered $\beta$-phase in the region adjacent to the melting curve in the entire pressure range studied by us.  It can be deduced not just from the presence of the step-like anomaly below the melting transition on all curves in Fig.~\ref{eps} but also from specific dispersion of dielectric response in the radio-frequency range intrinsic to the $\beta$-phase which was discussed earlier.  Also the data shown in Fig.~\ref{ag} demonstrate very distinct shoulders on the temperature dependence of static conductivity in the vicinity of $\beta-\gamma$ and $\beta-\alpha$ phase boundaries. In some way such extended pressure  region of $\beta$-phase stability contradicts the conclusions of Ref.~\onlinecite{gromnitskaya:jetpl04} where vanishing of $\beta$-phase at pressures above 1.6 GPa was conjectured on the ground of their ultrasound data. However no direct evidence (like demonstration of triple point) was presented there, so we believe that in fact there is no contradiction in our data with the previous one. 

It is also worth mentioning the evolution of dielectric susceptibility anomalies along $\alpha-\delta$, $\gamma-\delta$ and $\alpha-\gamma$ phase boundaries. Though in the first two cases the anomalies are represented by rather smooth peaks widened with the pressure increase (see curves $P=1.45,4.1,6.0$ GPa in Fig.~\ref{eps} and Fig.~\ref{ag} too) while $\alpha \to \gamma$ transition is always accompanied by very sharp ($\lambda$-like) feature that was regularly observed in the pressure range $3.4-3.7$ GPa. Since this feature is strongly dependent on pressure, we had certain difficulties in exact determination of $\alpha-\gamma$ phase boundary so it is approximately designated in Fig.~\ref{pt} as dashed line. The experimental curve ($P= 3.4$ GPa)  in Fig.~\ref{eps} demonstrates variation  of static dielectric susceptibility between 4 solid phases of methanol.

Another interesting region on the phase diagram of methanol is the range in the vicinity of 3.7 GPa on the melting curve (Fig.~\ref{pt}). There is significant deviation of experimental points from the smoothed line which in some way reflects the trend of $\beta - \alpha$  and $\beta - \gamma$ phase boundaries but shifted to higher temperatures. The smoothed lines (solid lines in Fig.~\ref{pt}) in the case of solid methanol phases comply to the thermodynamical phase contact rule which leaves little freedom how the phase boundaries could be placed.  On the other hand, the ``wiggling'' of the melting curve as suggested by experimental data in Fig.~\ref{pt} would indicate existence of different solid phases below it (for example two $\beta-\beta'$ phases in our case). However no difference between $\beta$-phase in low-pressure region (adjacent to $\alpha$-phase) and high-pressure (adjacent to $\gamma$-phase) was found. The most plausible explanation is the certain arbitrariness in the way we have drawn the melting curve -- at the end of transition, where $\varepsilon$ is equal at the beginning and the end of the cooling/heating cycles, but not in the middle of melting process, which is much more accepted practice. For example, one may expect that  the width of melting transition might be widened in the vicinity of $\alpha-\gamma$ transition, but this certainly requires a more thorough investigation. 

Nonetheless our measurements can be considered not just as an independent experimental evidence of existence of high-pressure methanol $\gamma$-phase\cite{allan:prb98} at room temperature, but also as a strong indication of existence of another low-temperature high-pressure $\delta$-phase. 
\section{Vitrification of methanol}
\begin{figure}  
\centering  
\includegraphics[width=0.8\columnwidth]{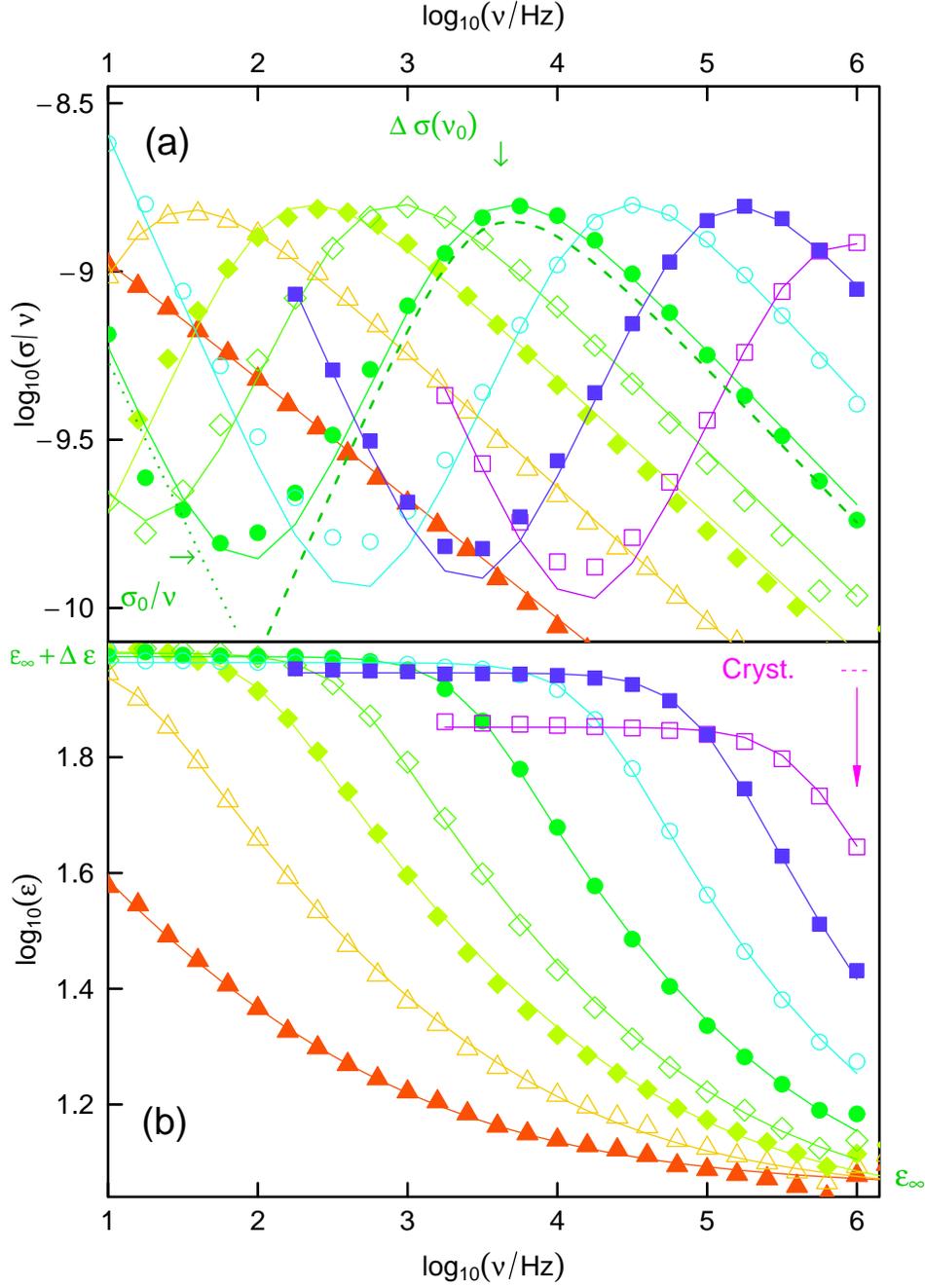} 
\caption{Imaginary (a) and real (b) parts of complex dielectric susceptibility of glassy methanol measured in the process of heating at pressure $P=4.3$ GPa and different temperatures: $\square$ -- 250.5, $\blacksquare$ -- 246.4. $\circ$ -- 236.5, $\bullet$ -- 219.7, $\lozenge$ -- 213.6, $\blacklozenge$ -- 210.6, $\triangle$ -- 205.4, $\blacktriangle$ -- 200.4 K. The symbols are experimental points, the solid lines are fits according to Eq.~\ref{cd}. Schematic drawing (dashed and dotted lines) on graphics illustrate the meaning of fitting parameters in Eq.~\ref{cd}. }
\label{glass}  
\end{figure}

Quite unexpectedly  we obtained methanol in glassy state (Fig.~\ref{pt}) in the narrow pressure range $4.3-4.5$ GPa. The process of vitrification is illustrated in Fig.~\ref{glass}   by evolution of structural relaxation in supercooled liquid methanol at pressure $P=4.3$ GPa and temperature range  $T=150-250$ K.  Though relaxation properties of many mono-alcohol liquids have quite complicated character, involving several components of mixed Debye and non-Debye types, in our case we have observed only one non-Debye mode ($\beta \ne 1$ in Eq.~\ref{cd} below) which was fitted by the Cole-Davidson formula \cite{davidson:jcp51} with additional direct current contribution (solid lines in Fig.~\ref{glass}):
\begin{equation} 
\varepsilon(\nu)=\varepsilon_{\infty}+\frac{i \sigma_0}{\nu} + \frac{\Delta \varepsilon + i \Delta \sigma}{(1+i \nu/\nu_0)^{\beta}}
\label{cd}
\end{equation}

\begin{figure}   
\centering   
\includegraphics[width=0.8\columnwidth]{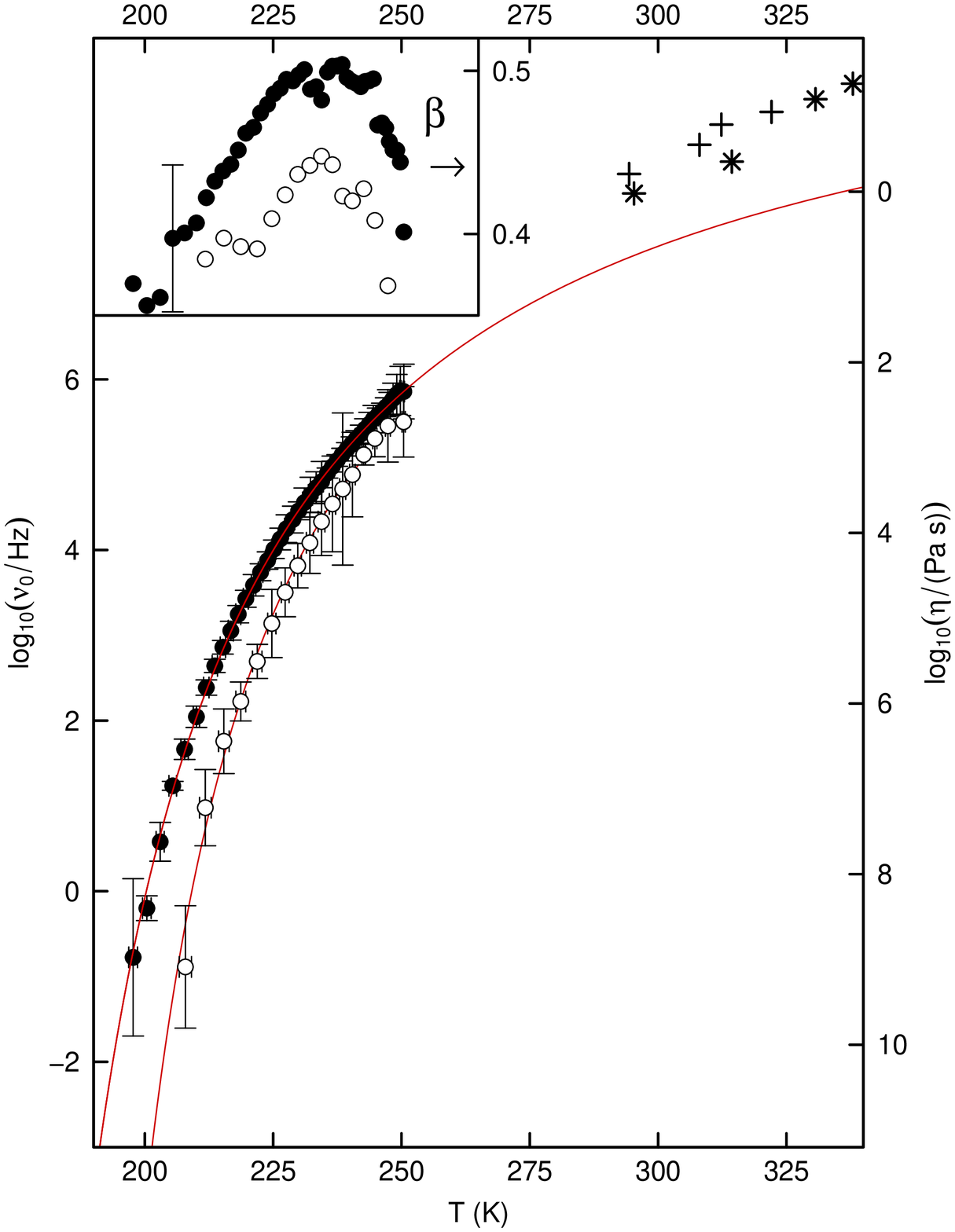} 
\caption{Comparison of temperature evolution of characteristic frequency $\nu_0$  from Eq.~\ref{cd} ($\bullet$ -- $P=4.3$ GPa, $\circ$ -- $P=4.6$ GPa) and viscosity data from Ref.~\onlinecite{grocholski:jcp05} ($+$ -- $P=4.1$ GPa, $*$ -- $P=4.6$ GPa). The left and right axis are related to one another according to Eq.~(\ref{maxwell}) with $G_\infty= 1$ GPa and $\tau=1/(2 \pi \nu_0)$. The thin lines on main plot are fits according to Eq.~\ref{vft}. Temperature evolution of parameter $\beta$ is shown in the inset.}   
\label{pars}   
\end{figure}

Characteristic frequency of relaxation ($\nu_0$)  is  shown in Fig.~\ref{pars}. Temperature evolution of $\nu_0$ along two isobars is approximated by empirical Vogel-Fulcher-Tamman (VFT) relation:
\begin{equation}
\log_{10}(\nu_0) \sim  \frac{T}{T-T_0}
\label{vft}
\end{equation}

By extrapolation of experimental data with this formula  the vitrification temperatures of methanol at $P=4.3-4.6$ GPa can be obtained as the temperature values at which $\nu_0=10^{-3}$ Hz ($T_g$= 192 and 202 K respectively). Fragility index  $m_p$, demonstrating how far the relaxation in real liquid is from the Arrhenius one \cite{angell-ngai:jap00}, yields the value about 85, i.e. methanol is more fragile  than the popular molecular glassformer glycerol \cite{pronin:pre10,pronin:jetpl10} at similar pressures. 

However, methanol glass obtained in this pressure range is quite unstable and easily crystallizes upon heating. Beginning of the crystallization process is visible in Fig.~\ref{glass} as a quite large drop of dielectric susceptibility at $T=250.5$ K, which at even higher temperatures (not depicted here) would bring it to the level of several units (that is on the same level with $\varepsilon_{\infty}$ in Fig.~\ref{glass}). It is interesting to note that this recrystallization process roughly coincides with $\gamma-\delta$ phase boundary, so the crystal obtained is $\gamma$-phase, which at higher temperatures converts into $\beta$-phase in accordance with the phase diagram in Fig.~\ref{pt}.
%\subsection{Discussion}

Some parameters of vitrification process can be compared with the literature data. Though (as it was already mentioned) methanol is hard to obtain in glassy form at ambient pressure, it is well known that addition of small amount of water ($> 6.5$ mol. \% \cite{bermejo:pla90}) expedites this process significantly. However, the DS data of vitrification in methanol-water mixtures have only recently became available\cite{sun:jpcb11}. In small concentrations (10 and 20 mol. \%) the relaxation process clearly consists of two seemingly non-symmetric (and consequently non-Debye) modes, which were ascribed to methanol and water. By fitting methanol component relaxation time published there and using the same conventions as above we obtain values $T_g=109-114$ K (depending on water content) and $m_p$ = 65,  which are in good accordance with our results. Relatively slow rise of glassification temperature with pressure rising and increase of fragility at high pressure was previously registered in many hydrogen-bonded glassformers (e.g. glycerol\cite{pronin:pre10,pronin:jetpl10}). 

Glassy methanol obtained by compression is more controversial topic. First it was introduced by Piermarini\cite{piermarini:jap73} as an explanation of widening of ruby  R$_1$  fluorescence line observed in diamond anvils filled with pure methanol at 8.6 GPa. However, the glassification pressure $P_g=$ 8.6 GPa is almost certainly an error (though it is still occasionally cited\cite{loubeyre:jcp13}), and most likely this widening  was caused by crystallization of methanol at this pressure. Moreover, it was directly disproved by viscosity measurements\cite{cook-herbst:jpc93} at similar conditions ($P=$8.3 GPa, $\eta \approx 10^4$ Pa~s, that is quite liquid-like) and analysis of methanol crystallization process at high pressures\cite{brugmans:jcp95}. Extrapolation of viscosity data \cite{cook-herbst:jpc93,grocholski:jcp05} suggests larger values $P_g = 11-20$ GPa (depending on the model used), which roughly corresponds to the results of Ref.~\onlinecite{brugmans:jcp95}. Though the translationally disordered methanol phase was observed at pressures above 10 GPa\cite{brugmans:jcp95,zaug:jpc94}, no other information on its nature and relaxation properties is yet available.

We can compare methanol's DS relaxation frequencies at $P=4.1-4.6$ GPa with viscosity data\cite{grocholski:jcp05} (see Fig.~\ref{pars}) obtained in the same pressure range, but at higher temperatures ($T=298-245$ K). Extrapolation of viscosity data by the Arrhenius law yields\cite{grocholski:jcp05} the values $T_g= 120 -160$ K, but this is surely is an underestimate because this extrapolation does not take into account crossover to  steeper dependence (like the VFT one) close to the glassification temperature. Viscosity ($\eta$) and relaxation time ($\tau$) are related to each other by the Maxwell relation:
\begin{equation}
\eta = \frac{\tau}{G_\infty}
\label{maxwell}
\end{equation}
where $G_\infty$ is an infinite frequency shear module. It is of order $\approx 50$ GPa for common window-pane glass, but for small molecule organic glassformers it is likely to be lower and lays in the range $0.9-9$ GPa (characteristic for two popular molecular glassformers DGEBA and glycerol respectively) \cite{schroter:jncs02}. For methanol we choose the value $G_\infty =$ 1 GPa which is comparable by order of magnitude with the value suggested by the viscosity data of its close analog -- ethanol\cite{II*stickel}. Though VFT fit of our data yields a slightly underestimated value of characteristic frequency than one would expect from viscosity data  (Fig.~\ref{pars}), this sort of discrepancy was already observed  in molecular galssformers (see the comparison of ``Maxwell'' and DS times in DGEBA in Ref.~\onlinecite{schroter:jncs02}).  There may be two sources of this discrepancy: either the contribution of another high-frequency process to the overall viscosity (as a result the substance is less viscous than it could be expected from the consideration of only one low-frequency relaxation), or, most probably, the dynamic crossover observed in most of organic molecular glassformers\cite{II*stickel}. For example the extrapolation of low temperature data in ethanol at ambient pressure by the VFT relation produces underestimated values of characteristic frequency \cite{lunkenheimer-kastner:pre10}.

Still our measurements are the first parametrization of glass transition in methanol and the first report of vitrification of methanol by cooling at high pressure.
\section{Discussion}
The most interesting question arising from our measurements is if there is relation  between vitrification of methanol and phase transitions taking place in adjacent pressure ranges. Answering this requires  closer examination of diffraction data  available from literature \cite{tauer:ac52,krishna:ijp59,torrie:mp89,torrie:jssc02}. The main point of this examination is establishing the type of phase transformations observed in methanol at high pressures. Indeed the ordering of hydrogen bonds in solid methanol at lower temperatures can be formally described by a symmetry loss. This sort of transformations is described as displacive or order/disorder ones which as a rule bring about quite small energy output in the transition. On the other hand, another type (called reconstructive) of  phase transitions accompanied with formation of new bonds (and new symmetry operations, not present in high or low temperature phases) will produce a much higher energy output (see e.g. Refs.~\onlinecite{en*izyumov82,toledano:90,toledano:96}). Though the energy of hydrogen bonds present in molecular methanol isn't large, but one may expect that reconstructive phase transitions in solid methanol (for example in $\alpha - \beta$ transformation) would result in much greater enthalpy output than registered in practice \cite{staveley:jcs54}. It was already shown that this output (as suggested by Ref.~\onlinecite{wurflinger:jpcs77}) as well as the volume effect\cite{gromnitskaya:jetpl04}  diminishes with pressure rising, so one may assume that the phase transformations in methanol (at least $\alpha - \beta$ one) are displacive. Thus a certain restrictions may be applied to the experimental structural data reviewed below.  

 Though all the authors \cite{tauer:ac52,torrie:mp89,torrie:jssc02} are unanimous about the structure of $\beta$ phase (space group $C m c m$ with 4 molecules per conventional unit cell $Z=4$), there is disagreement about the $\alpha$-phase. The controversy can be summarized as whether $\beta \to \alpha$ transformation leads to multiplication of the unit cell or loss of inversion center in the $\alpha$-phase. The first X-Ray measurements of single-crystal \cite{tauer:ac52} and polycrystalline \cite{krishna:ijp59} samples demonstrated that the center of inversion is retained (space group $P2_1/m$), but the conclusion about the cell size was not that certain. The value $Z=2$  was suggested as a preferred one but allowances \cite{tauer:ac52}  were made for larger cell which  is twice the original primitive unit cell with duplication of $a$ parameter. Next 30 years of optical \cite{falk:1554,wong:jcp71,dempster:jcp71,durig:jcp71,franck:fdcs78,mammone:jpc80,anderson:jrs88}  and NMR \cite{garg:jcp73} research demonstrated that the number of optical modes is only compatible with prepositions that either $Z > 2$, or there is no inversion center, or both. The structure version with orthorhombic lattice ($P2_12_12_1$; $Z=4$) suggested lately \cite{torrie:mp89,weng:pss92,torrie:jssc02} realizes just that last case. Still it can be easily demonstrated that in this case $\beta \to \alpha$  phase transition is reconstructive one. So the earlier monoclinic version but with larger cell ($P2_1/m$; $Z=4$) seems more preferable for the $\alpha$-phase. 

Such duplication of unit cell can be only produced by lattice distortions with wave vector in $S$ ($k=(1/2,1/2,0)$) point of Brillouin-zone of $\beta$-phase. Moreover, all other observed transformations in methanol at high pressure can be explained in similar way as freezing of vibrational modes in the $S$ point. Using available software\cite{isotropy,capillas:jac03,aroyo:zfk06} one can demonstrate that any of the four 2-dimensional irreducible representations in the $S$-point of $C m c m$ space group involves not only the symmetry breaking with the full 2-dimensional order parameter ($\eta$ in Fig.~\ref{pt} and Table~\ref{struct}) but also has two more symmetrical directions (isotropical) in the order parameter space. The identification of methanol phases and unit cell transformations brought about by $S_2^-$ irreducible representation of $C m c m$ space group is shown in Table~\ref{struct}. 

\begin{table*}[h]
\caption{ Previously determined solid methanol phases with respective references (column 1) and structural parameters of solid methanol, number of molecule per unit cell $Z$ and  molecular volume $V'$ (columns 2-3). The corresponding structures induced  by the  $S_2^-$ irreducible representation of $C m c m $ space group with respective isotropy order parameter $\eta$  and approximate transformations relating the hypothetical conventional unit cell ($a,b,c$) to the $\beta$-phase one ($a',b',c'$) are shown in columns 4-5.}
\label{struct}
\centering
\begin{tabular}{lllll}
\toprule%
Phase(Refs.)                                      & S.G./Z/$V'$(\AA$^3$)& Unit cell(\AA, $^{o}$) & S.G./Z($S_2^-$) &Transformation \\\hline%  
$\beta$\cite{tauer:ac52,torrie:mp89,torrie:jssc02} &  $C m c m$    &       $a=6.41$    & $Cmcm$        &                  \\
                                                   &  $Z=4$        &       $b=7.20$    & $Z=4$          &                  \\
                                                   &  $V'=53.0$    &       $c=4.64$    & $\eta=(0,0)$   &                  \\\hline
$\alpha$\cite{tauer:ac52,krishna:ijp59}            &  $P2_1/m$     &    $a=4.53$ (4.59) &   $P2_1/m$     & $a=a'+b'$       \\
                                                   &  $Z=2$        &    $b=4.69$ (4.68) &   $Z=4$       & $b=a'/2-b'/2$     \\
                                                   &  $V'=52.4$    &    $c=4.91$ (4.92) &$\eta=(\alpha,0)$  & $c=c'$        \\
                                                   &               &  $\gamma=90\pm 3$ (97.5)&          &     \\\cline{2-3}
$\alpha$\cite{torrie:mp89,torrie:jssc02}           &  $P2_12_12_1$  &    $a=8.87$        &               &                  \\
                                                   &  $Z=4$        &    $b=4.64$        &               &                   \\
                                                   &  $V'=52.1$    &    $c=4.87$        &               &                   \\\hline
$\gamma$\cite{allan:prb98}                       & $P\overline{1}$ &    $a=7.67$        &   $C m c m$   &   $a=2a'$        \\
                                                   &  $Z=6$        &    $b=7.12$        &   $Z=16$      &   $b=2b'$        \\
                                                   &  $V'=39.5$    &    $c=4.41$        &$\eta=(\alpha,\alpha)$&   $c=c'$   \\
                                                   &               &    $\alpha=$ 88.10 &               &                  \\
                                                   &               &    $\beta=$ 93.85  &               &                  \\
                                                   &               &    $\gamma=$102.2  &               &                  \\\hline
$\delta$                                           &               &                    &    $P2_1/m$   &    $a=a'+b'$     \\
                                                   &               &                    &    $Z=8$      &    $b=a'-b'$     \\
                                                   &               &                    &$\eta=(\alpha,\beta)$ &    $c=c'$  \\\toprule
\end{tabular}
\end{table*}

This consideration is somewhat oversimplified and does not take into account possible coupling of $S$- and $\Gamma$-point order parameters which would lead to even higher symmetry breaking in $\gamma$-phase down to $P\overline{1}$ space group observed earlier in $\gamma$-phase \cite{allan:prb98}.
 However the center of inversion is retained in transitions $C m c m \to P2_1/m$ (or $C m c m \to P\overline{1}$) and it can be shown \cite{toledano:prb80,isotropy} that these transformations are improper ferroelastic ones. Though in general the  spontaneous elastic strain in improper ferroelastic transitions is not large, it can be observed in samples subjected to external mechanical fields, for example sound waves. In methanol, for example, the volume effect in $\beta-\alpha$ transition at pressures $P < 1 $ GPa is small and the volume variation in the transition from liquid to $\alpha$-phase is quite smooth \cite{gromnitskaya:jetpl04}. However the transition $\beta \to \alpha$ as measured by ultrasound methods produces quite large variation in sound velocity values, so it may be an evidence of ferroelastic nature of this phase transition.

Moreover the relation between ferroelastic transitions and amorphization at high non-hydrostatic pressure was considered before \cite{toledano:prb05,en*braginsky:ftt90a,brazhkin:prl03} and this may be the rationale behind formation of the glassy state in methanol at the pressures $P=$4.3, 4.6 GPa. Roughly speaking the process responsible for amorphization of methanol may result from the  hindering of nucleus growth because of elastic strain between domains of different ferroelastic phases ($\alpha$,$\gamma$,$\delta$) in the vicinity of phase boundaries between them. The same consideration is applicable to the methanol glass obtained by fast compression at room temperature \cite{brugmans:jcp95}. The route of this process also lays in the vicinity  (but at higher temperatures) of this region. So it probably involves quenching of liquid methanol into $\delta$-phase, which is likely to be present at room temperature at higher pressures ($\approx 10$ GPa) as suggested by the phase diagram (Fig.~\ref{pt}). 

The straightforward test of the structural model of phase transformations in methanol is determining the density of $\gamma$-phase. Experimental value \cite{allan:prb98} $Z=6$ was based on the density value measured in liquid methanol \cite{brown:s88}, which is likely to yield underestimated value of solid methanol density. In the present treatment the primitive unit cell with comparable volume should contain $Z=8$ methanol molecules, so it should be at least 25 \% denser. But the resolution of this contradiction requires a more thorough examination of methanol solid phases  structures at  ambient and high pressures in wide temperature range.   
\section{Conclusions}    
Dielectric spectroscopy measurements in methanol in the pressure range up to $P=6.0$ GPa demonstrate the existence of two high-pressure phases of methanol. The room-temperature one obviously corresponds to the known $\gamma$-phase \cite{allan:prb98}, and  the lower-temperature one is a previously unknown phase (tentatively called $\delta$-phase). In the intermediate pressure region  $P=4.3-4.6$ GPa we observed vitrification of methanol and evaluated phenomenological parameters describing its temperature evolution. We suggest a simple structural model describing phase transformations in methanol as condensation  of vibrational modes in $S$-point of the Brillouin zone of disordered $\beta$-phase. Possible relation of this model with vitrification of methanol at high pressure was considered too. 
\begin{acknowledgements}
This work was supported
 by the RFBR grants \#13-02-00542 and \#13-02-01207. The authors are grateful for A.V. Rudnev and A.V. Gulutin for technical assistance in accomplishing of experiments.
\end{acknowledgements}
%\bibliography{meoh_a}
%\input{m_a.bbl}


\end{document}